\documentclass[showpacs,aps,preprint,graphicx,psfig]{revtex4}
\usepackage{amsmath}
\usepackage{graphicx}

\input epsf.sty
\input{tcilatex}

\begin{document}

\title{Obvious enhancement of the total reaction cross sections for $%
^{27,28} $P with $^{28}$Si target and the possible relavent mechanisms}
\author{Z.H. Liu, M. Ruan, Y.L. Zhao, H.Q. Zhang, F. Yang, Z.Y. Ma, C.J.
Lin, B.Q. Chen, Y.W. Wu}
\affiliation{China Institute of Atomic Energy, Beijing 102413, China}
\author{W.L. Zhan, Z.Y. Guo, G.Q. Xiao, H.S. Xu, Z.Y. Sun, J.X. Li, Z.J. Chen%
}
\affiliation{Institute of Modern Physics, Chinese Academy of Science, Lanzhou 730000,
China}

\begin{abstract}
The reaction cross sections of $^{27,28}$P and the corresponding isotones on
Si target were measured at intermediate energies. The measured reaction
cross sections of the $N=12$ and $13$ isotones show an abrupt increase at $%
Z=15$. The experimental results for the isotones with $Z\leq 14$ as well as $%
^{28}$P can be well described by the modified Glauber theory of the optical
limit approach. The enhancement of the reaction cross section for $^{28}$P
could be explained in the modified Glauber theory with an enlarged core.
Theoretical analysis with the modified Glauber theory of the optical limit
and few-body approaches underpredicted the experimental data of $^{27}$P.
Our theoretical analysis shows that an enlarged core together with proton
halo are probably the mechanism responsible for the enhancement of the cross
sections for the reaction of $^{27}$P+$^{28}$Si.
\end{abstract}

\date{December 23,2003}
\maketitle

I. INTRODUCTION

Recently, Navin et al \cite{refNa98} have confirmed the important role of
the $\pi s_{1/2}$ orbital in the predicted halo structure \cite%
{refBr96,refRe96,refCh98} of the neutron-deficient phosphorus isotopes $%
^{26,27,28}$P by measurements of deexcitation $\gamma $ ray in coincidence
with the momentum distribution of the projectile residues. However, the
measurements \cite{refZh02} of reaction cross sections for $^{27,28}$P+$%
^{12} $C at intermediate energies do not show proton-halo structure in $%
^{28} $P. Generally speaking, large halos are possible only for
valence-neutrons in the s- and p-states, and the effect of the Coulomb
barrier would hinder the formation of a proton halo \cite{refRi92}. Hence,
proton halos are more difficult to probe experimentally, and the conclusions
extracted may be not clear cut. $^{8}$B is a typical example. Many
experiments \cite{refMi92,refBl97,refNe96,refKe96}\ have been devoted to
studies of $^{8}$B in order to establish its halo nature. Although
investigated in considerable details, its halo character has been in
controversy till recent years. Similar situation may occur in $^{28}$P.
Thus, it is an interesting problem whether the proton halo structure really
exists in $^{28}$P or how large the halo is if existed. Moreover, recent
studies of nitrogen, oxygen, fluorine isotopes show an abrupt rise in
interaction cross section $\left( \sigma _{I}\right) $ at N=15 \cite{refOz01}
which are underpredicted even with 100\% s-wave probability of a valence
neutron in a `` core-plus-neutron'' halo model \cite{refOz01,refKa01}. It is
proposed that a core modification takes place in these nuclei \cite{refKa01}%
.Kanungo et al \cite{refKa02} measured the longitudinal momentum
distributions of one- and two-neutron removal fragments ($^{21,22}$O) of $%
^{23}$O from the reaction with a carbon target at 72 MeV/nucleon. Their
results indicate the modification of core ($^{22}$O) structure for the $sd$
shell nuclei near the neutron-drip line. The present work is motivated by
the observations of this new type of an anomaly in $sd$ shell nuclei. For
this purpose, reaction cross sections of isotonic nuclei with N=12 and 13 on
Si target were measured, and special attentions were payed to the nuclei
with Z=15, i.e., $^{27,28}$P.

\medskip

II. EXPERIMENTAL PROCEDURE AND\ RESULTS

The experiment was performed at the Institute of Modern Physics, Lanzhou.
Secondary beams of $^{27,28}$P and the corresponding isotones were produced
by the projectile fragmentation of an $^{36}$Ar primary beam on a Be
production target at 69 MeV/nucleon. The Be production target was 98.8 mg/cm$%
^{2}$ in thickness. The isotopes of the secondary beams were separated and
selected by the magnetic rigidity of Radioactive Ion Beam Line in Lanzhou
(RIBLL) \cite{refZh02} which served as a double-achromatic magnetic
spectrometer in the present experiment. An Al energy degrader was used to
improve the momentum resolution and purity of the secondary beams. The time
of flight (TOF) of the projectiles was determined by two scintillators
placed at the first and second achromatic focal planes of RIBLL 16.8 m
apart. The resolution of TOF was 4 ns. The position information was given by
two parallel-plate-avalanche countors (PPACs) placed in the front of and
behind the second scintillator. Finally, a telescope consisting of seven
transmission Si surface barrier detectors was installed after the second
PPAC. The thicknesses of these detectors were 150 $\mu m$ for the first one
and 300 $\mu m$ for the others. The TOF information along with the energy
deposition ($\Delta E_{i}$) in the Si detectors were used to identify those
projectiles of interest which underwent reactions. Fig. 1 illustrates a
typical two-dimensional plot of TOF versus $\Delta E_{2\text{. }}$It is seen
from the figure that particles can be identified clearly by using TOF and $%
\Delta E_{i}$. Apart from $\Delta E$ detection, some of the Si detectors
also served as the reaction target. Hence, the use of multiple Si detectors
permits simultaneous measurement of reaction cross sections $\left( \sigma
_{R}\right) $ for several different energies.

Our data analysis procedure is similar to that used by Warner et al \cite%
{refWa95,refWa98}. A tight gate on PPACs, TOF and $\Delta E_{i}$ was set for
each detector to identify projectiles which had not yet reacted in that and
preceding detectors. Fig. 2 shows a spectrum of the total energy deposited
in the telescope by $^{28}$P projectiles. Events left the dotted line were
counted as reaction ones. The probability $\eta _{1}$ for a reaction to
occur beyond the first Si detector was determined by the ratio of the
reaction events to the total events in the spectrum which was gated on
PPACs, TOF and $\Delta E_{1}$. Likewise, the probability $\eta _{i+1}$ for a
reaction to take place beyond the $\left( i+1\right) st$ detector was found
from a total energy spectrum gated on PPACs, TOF and $\Delta Es$ of the $%
\left( i+1\right) st$ and all preceding $Si$ detectors. From the measured $%
\eta _{i}$ and $\eta _{i+1}$, the average $\sigma _{R}$ corresponding to the
reactions taken place in the $i$th Si detector was determined by

\begin{equation}
\sigma _{R}=\frac{A}{\nu \rho \left( \Delta x\right) _{i}}\ln \left[ \frac{%
1-\eta _{i+1}}{1-\eta _{i}}\right] ,  \label{eq1}
\end{equation}%
here $A$ and $\rho $ are atomic mass number and density of target, $\nu $ is
Avogadro's number, and $\left( \Delta x\right) _{i}$ the thickness of $%
\Delta E_{i}$. The $\sigma _{R}$ was corrected for the reaction events under
the elastic peak by extrapolating the spectrum left the dotted line. This
correction only accounts for a few percentages of the total reaction cross
section. The error in $\sigma _{R}$ includes the statistics, uncertainties
of the detector thickness and the extrapolation of reaction events. The
measured reaction cross sections are listed in Table 1.

Fig.3 shows the measured $\sigma _{R}$ (solid squares) as a function of Z
for isotones with N=12 and 13 at 40 MeV/nucleon. It is worth to note that $%
\sigma _{R}$ increases obviously at Z=15. The situation is very similar to
the nitrogen, oxygen, fluorine isotopes where a large increase in $\sigma
_{I} $ at N=15 was observed \cite{refOz01,refKa01}. This similarity may be a
signature of charge-independence of nuclear force in the nuclei far from $%
\beta $-stability line. In addition, it is seen that the rise of cross
section for $^{27}$P (even N case) is much more abrupt than that for $^{28}$%
P (odd N case). Again, this feature is very similar to the nitrogen, oxygen,
fluorine isotopes \cite{refKa01}. In the latter case, for the even Z nuclei (%
$^{11}$Be, $^{19}$C, $^{23}$O) the rise of cross section is rather abrupt,
however for the odd Z nuclei ($^{22}$N, $^{24}$F) the cross section shows a
continuously increasing trend. These even-odd features are probably a
reflection of the effect of pairing interaction \cite{refKa01}.

\bigskip

III. MODIFIED GLAUBER MODEL ANALYSIS

A halo nucleus is considered to be composed of a core with one or two
loosely bound nucleons tunneling out at distances far away from the core %
\cite{refJe00}. An abrupt enhancement of cross section of a nucleus compared
to its preceding isotope/isotone neighbours can be a signature of a halo
structure. The structure of halos is usually analyzed by the "core-plus-halo
nucleon(s)" model \cite{refKa01,refKa02}, which is realized with a few-body
(FB) Glauber model \cite{refOg92,refAl96}. In the FB Glauber model, the
projectile is decomposed into a core and halo nucleons, and the spatial
correlation between core, halo nucleons and target are explicitly taken into
account \cite{refOg92,refAl96,refZh03}. When the nucleus has only one halo
nucleon, the reaction cross section is given by,

\begin{equation}
\sigma _{R}^{FB}=\int d\mathbf{b}\left\{ 1-\left| \left\langle \varphi
_{0}\left| \exp \left[ i\chi _{FT}\left( \mathbf{\bar{a}}\right) +i\chi
_{nT}\left( \mathbf{\bar{a}}+\mathbf{s}_{1}\right) \right] \right| \varphi
_{0}\right\rangle \right| ^{2}\right\} \text{,}  \label{eq2}
\end{equation}

\begin{equation}
i\chi _{FT}\left( \mathbf{\bar{a}}\right) =-\int d\mathbf{s}T_{F}\left(
\mathbf{s}\right) \int d\mathbf{t}T_{T}\left( \mathbf{t}\right) \Gamma
\left( \mathbf{\bar{a}}+\mathbf{s}-\mathbf{t}\right) \text{,}  \label{eq3}
\end{equation}

\begin{equation}
i\chi _{nT}\left( \mathbf{\bar{a}}+\mathbf{s}_{1}\right) =-\int d\mathbf{t}%
T_{T}\left( \mathbf{t}\right) \Gamma \left( \mathbf{\bar{a}}+\mathbf{s}_{1}-%
\mathbf{t}\right) \text{,}  \label{eq4}
\end{equation}%
where $\mathbf{b}$ is a two dimension vector of the impact parameter which
is perpendicular to the incident direction, $\mathbf{\bar{a}}=\mathbf{b}-%
\frac{1}{A}\mathbf{s}_{1}$ is the impact parameter vector of core, $A$ is
the mass number of the projectile, $\mathbf{s}_{1}$ is the perpendicular
component of the halo nucleon coordinate with respect to the mass center of
the core, and $\varphi _{0}$ is the bound state wave function. $\chi _{FT}$,
$\chi _{nT}$ are the optical phase-shift functions of the core and halo
nucleon scattering with the target, respectively. $T_{F}$ is the thickness
function of the core. $\Gamma $ is the profile function of a nucleon-nucleon
$\left( N-N\right) $ scattering. In our calculations, it takes the following
expression \cite{refCh97},

\begin{equation}
\Gamma \left( \mathbf{b}\right) =\frac{\sigma _{NN}}{2\pi \beta ^{2}}\left(
1-i\alpha _{NN}\right) \exp \left( -\frac{b^{2}}{\beta ^{2}}\right) \text{,}
\label{eq5}
\end{equation}%
where $\sigma _{NN}$ is the total $N-N$ scattering cross section, $\alpha
_{NN}$ is the ratio of the real to the imaginary part of the forward $N-N$
scattering amplitude, and $\beta $ represents for the finite range of the $%
N-N$ interaction, respectively. It is important to take the finite range of
the $N-N$ interaction into account in order to reproduce the experimental
data at low and intermediate energies \cite{refZh01a}. The range of the $N-N$
interaction is fixed as $\beta =1.0$ fm in the present work. The large
enhancement of the experimental reaction cross sections of $^{27,28}P$ calls
for careful analysis with the Glauber theory of OL and FB approaches
described above.

In Fig.3, the predictions of the modified Glauber theory in the optical
limit (OL) approach\cite{refZh01a,refZh01b} are compared with the
experimental cross sections. In this approach, the Coulomb and finite range
corrections are taken into account. It is verified \cite{refZh01a,refZh01b}
that with this modified version the Glauber theory can be extended to low
energy region. In our calculations, the nuclear density distributions are
evaluated in a Woods-Saxon (WS) potential. The radius and diffuseness
parameters are taken as $r_{0}=1.17$ fm and $a=0.65$ fm. The depth of the WS
potential is adjusted by reproducing the single proton separation energy.
The proton separation energies for these isotones are also listed in Table
1. There is only one exception of $^{25}$Al. Small separation energy of the
valence proton in $^{25}$Al results in too diffused density. To reproduce
the experimental data, the neutron separation energy is used to adjust the
WS potential depth in the calculation for $^{25}$Al. It may be seen from
Fig.3 that there is satisfactory agreement between theory and experiment for
the isotones with $Z\leq 14$. The experimental datum for $^{27}$P appears to
be obviously greater than the calculated value. Although the reaction cross
section of $^{28}$P displays an enhancement in comparison with the neighbor
isotone, the modified Glauber theory with a diffused density distribution of
$^{28}$P could describe the experimental datum.

As shown in Fig.4 the measured $\sigma _{R}$ of $^{27}$Si+$^{27}$Si can be
well described by the modified Glauber theory of OL approach. In the
calculations, the geometry parameters of the WS potential are fixed at the
values of $r_{0}=1.17$ fm and $a=0.65$ fm, and the depths of the WS
potential are adjusted by reproducing the valence proton separation energies
of $^{27}$Si and $^{28}$P, respectively. Due to the depths of the WS
potential are different, the root-mean-square (rms) radii of the bare and
core nuclei $^{27}$Si do not have the same values. The calculated rms radii
are 2.854 fm and 2.997 fm for the bare and core nuclei $^{27}$Si,
respectively. This means that the size of the core $^{27}$Si in the nucleus $%
^{28}$P is enlarged by about 0.143 fm as compared to that of the bare
nucleus $^{27}$Si. Due to the Coulomb barrier the rms radius of the proton
in the $2s_{1/2}$\thinspace state of $^{28}$P is only $%
<r_{h}^{2}>^{1/2}=4.016$ fm in the WS geometry $\left( r_{0},a\right)
=\left( 1.17,0.65\right) $ fm. Therefore, the enhancement of the measured $%
\sigma _{R}$ of $^{28}$P+$^{28}$Si could be described satisfactorily by the
size enlargement of the core $^{27}$Si and the wave function of the valence
proton at $2s_{1/2}$\thinspace state in the modified Glauber theory of the
OL approach. It should be pointed out that the theoretical result
underestimates the reaction cross section of $^{28}$P when the density
distribution of $^{28}$P is calculated in the nonlinear relativistic mean
field theory (RMF), where the density distribution of the valence proton at $%
2s_{1/2}$\thinspace state is less diffused.

The nucleus $^{26}$Si is an isotope with two neutron deficit. As shown in
Fig.5, the Glauber theory of the OL approach gives a well description of the
$^{26}$Si experimental data if a diffused density distribution with the WS
geometry of $\left( r_{0},a\right) $ $=\left( 1.27,0.9\right) $ fm is used.
The obtained rms radius of the bare nucleus $^{26}$Si is $3.190$ fm. Adding
one proton in the $2s_{1/2}$ state and adjusting the depth of the WS
potential to fit the separation energy $S_{p}=0.900$ MeV of the valence
proton, with the same geometry parameters the density distributions of the
core $^{26}$Si and valence proton are calculated. In terms of these density
distributions, the cross sections for the reaction of $^{27}$P+$^{28}$Si are
evaluated in the Glauber theory of the OL and FB approaches, respectively.
In this calculation, the valence proton at $2s_{1/2}$\thinspace state has a
relative diffused density distribution due to its weak binding energy, and
therefore the core is enlarged. The rms radii of the core and valence proton
extracted from these density distributions are $<r_{c}^{2}>^{1/2}=3.470$ fm
and $<r_{h}^{2}>^{1/2}=4.875$ fm, respectively. The difference between the
rms radii of the core and bare nuclei $^{26}$Si is $0.280$ fm. Even though
the results still underpredict the experimental data, which are shown in
Fig.5 as open diamonds and squares, respectively. In order to improve the
agreement, we increase the WS potential diffuseness of the valence proton to
$a=1.1$ fm, meanwhile keep the radius parameter and the core density
distribution fixed. In this way, the reaction cross sections of $^{27}$P+$%
^{28}$Si are recalculated with the Glauber model of the OL and FB
approaches. The resulting reaction cross sections are increased slightly,
and are still lower than the experimental data. In this case, the rms radius
of the valence proton is $<r_{h}^{2}>^{1/2}=5.235$ fm. On the other side,
the failure to reproduce the $^{27}$P data in detail may reflect
deficiencies in our treatment of the reaction cross section. For example,
the role of Coulomb-induced reaction is not taken into account in the
present modified Glauber model. However, in the cases of $^{28}$P and the
isotones with $Z\leq 14$ the modified Glauber theory could well describe the
experimental cross sections. Therefore, the contribution of the
Coulomb-induced reactions to the total reaction cross section, if any, may
be not important for the system of $^{27}$P+$^{28}$Si as well.

$^{27}$Mg and $^{28}$Al are the mirror nuclei of $^{27,28}$P, respectively.
Because of isospin symmetry, the level structures within each pair should be
similar. Therefore, it would be very interesting to make a comparison
between the mirror nuclei. Listed in Table 2 are the interaction cross
section ($\sigma _{I}$) for $^{27}$Mg+$^{12}$C \cite{refSu98} and reaction
cross section for $^{28}$Al+$^{12}$C \cite{refCa02} along with the results
of Glauber model analysis. In these calculations, the nuclear density
distributions are evaluated in the RMF theory \cite{refHo81,refRe86,refPa93}
with the parameter NL3 \cite{refLa97}. We calculate the cross sections by
the Glauber theory of the OL approach with and without finite range
correction. It can be seen from Table 2 that the results of these
calculations are in good agreement with the experimental data for $^{27}$Mg,
$^{28}$Al, but not for $^{27,28}$P. In the case of $^{27}$P, the usual
Glauber theory, i.e., the theory without finite range correction,
underpredicts the experimental datum about 50\%. Therefore, the comparison
with the mirror nuclei supplies us a collateral evidence that the
proton-rich phosphorus isotopes $^{27,28}$P should have anomalous structures.

\bigskip

IV. SUMMARY

The reaction cross sections of $^{27,28}$P and the corresponding isotones on
Si target are measured at intermediate energies. The measured reaction cross
sections of the $N=12$ and $13$ isotones show a large increase at $Z=15$.
The experimental results for the isotones with $Z\leq 14$ as well as $^{28}$%
P can be well described by the modified Glauber theory of the OL approach.
The enhancement of the cross section for the $^{28}$P+$^{28}$Si reaction
could be well explained by the modified Glauber theory of the OL approach
with an enlarged core. The valence proton in $^{28}$P at $2s_{1/2}$%
\thinspace state has less diffused density distribution than usually
observed in a halo nucleon. The modified Glauber theory of the OL and FB
approaches somehow underpredicts the experimental data of $^{27}$P. In these
calculations 100\% occupancy of the valence-proton in the s-orbital is
assumed. Since s-wave contribution gives the largest cross section, the
results of the modified Glauber model calculation represent the up-limit
predictions of ``core-plus-halo nucleon(s)'' model. In addition, as shown in
Table 2 the modified Glauber model with the RMF theory densities also
underpredict the cross section for $^{27}$P. Although a satisfactory
agreement between the theoretical predictions and experimental data is not
reached, our theoretical analysis indicates that an enlarged core together
with proton halo are probably the mechanisms responsible for the anomalous
enhancement of the cross sections for the reaction of $^{27}$P+$^{28}$Si.
However, this suggestion should be viewed as a primary theoretical
explanation. Actually, the halo structure of $sd$ shell proton-rich nuclei
is not quite clearly understood theoretically yet. In order to prove the
possible relevant mechanisms, further investigations with more sophisticated
experiments and theories are required.

\begin{acknowledgments}
This work was supported by the National Natural Science Foundation of China
under Grants No. 10075080, 10175092, 10235030, 10275092, 10275094 and the
Major State Basic Research Development Programme under Grant No. G200007400.
\end{acknowledgments}

\medskip\ \thinspace \thinspace \thinspace
\begin{table}[tbp] \centering%
\caption{ The valence-proton separation energy, energy of projectiles, experimental reaction cross sections  and theoretical reaction cross sections calculated with the Glauber model of the OL approach for the
isotones with N=12 and 13. For $^{25}$Al instead of proton separation energy, the neutron energy is given.\label{table1}}%
\begin{tabular}[t]{ccccccc}
\hline
Nucleus & S$_{p}$ & E$_{in}$ & E$_{av}$ & E$_{out}$ & $\sigma _{R}^{\exp }$
& $\sigma _{R}^{OL}$ \\
& (MeV) & (MeV/nucleon) & (MeV/nucleon) & (MeV/nucleon) & (mb) & (mb) \\
\hline
$^{23}$Na & 8.794 & 19.3 & 14.5 & 9.7 & 1880$\pm $150 & - \\
&  & 22.7 & 22.7 & 19.3 & 2018$\pm $150 & 2010 \\
$^{24}$Mg & 11.693 & 22.4 & 17.7 & 13.1 & 1993$\pm $80 & - \\
&  & 29.4 & 25.9 & 22.4 & 1998$\pm $80 & 1963 \\
$^{25}$Mg & 12.064 & 27.4 & 23.9 & 20.3 & 2237$\pm $133 & - \\
&  & 33.4 & 30.4 & 27.4 & 2026$\pm $121 & 1967 \\
$^{25}$Al & 16.932$^{\ast }$ & 25.3 & 20.6 & 15.9 & 2141$\pm $120 & - \\
&  & 32.5 & 28.9 & 25.3 & 2027$\pm $110 & 1972 \\
$^{26}$Al & 6.307 & 30.4 & 26.7 & 23.0 & 2164$\pm $90 &  \\
&  & 36.6 & 33.5 & 30.4 & 2026$\pm $100 & 2075 \\
$^{26}$Si & 5.518 & 27.9 & 23.1 & 18.4 & 2351$\pm $190 & - \\
&  & 35.4 & 31.7 & 27.9 & 2284$\pm $190 & 2092 \\
$^{27}$Si & 7.463 & 33.0 & 29.2 & 25.3 & 2145$\pm $80 & - \\
&  & 39.5 & 36.3 & 33.0 & 2008$\pm $100 & 2050 \\
$^{27}$P & 0.900 & 30.6 & 25.8 & 20.9 & 3029$\pm $380 & - \\
&  & 38.4 & 34.5 & 30.6 & 2900$\pm $370 & 2302 \\
$^{28}$P & 2.066 & 35.6 & 31.6 & 27.6 & 2377$\pm $110 & - \\
&  & 42.4 & 39.0 & 35.6 & 2237$\pm $80 & 2210 \\ \hline
\end{tabular}%
\thinspace
\end{table}%

\medskip \medskip \medskip \medskip \thinspace \thinspace \thinspace
\begin{table}[tbp] \centering%
\caption{ A comparison between the mirror nuclei $^{27}$Mg,
$^{27}$P, and $^{28}$Al, $^{28}$P. The neutron separation energies
for $^{27}$Mg and $^{28}$Al, and proton separation energies for
 $^{27,28}$P are given in the Table. Listed in the sixth and seventh columns are the calculated
results of the Glauber model in the OL approach without and with
finite range correction, respectively, where the corresponding
density distributions are calculated in the RMF with the parameter
set NL3.
\label{table2}}%
\begin{tabular}[t]{ccccccc}
\hline
Nucleus & S$_{p}$ & reaction & E$_{Lab}$ & $\sigma ^{\exp }$ & ($\sigma
_{R}^{OL})_{\mathrm{nfc}}$ & ($\sigma _{R}^{OL}$)$_{\mathrm{fc}}$ \\
& (MeV) &  & (MeV/nucleon) & (mb) & (mb) & (mb) \\ \hline
$^{27}$Mg & 6.443 & $^{27}$Mg+$^{12}$C & 950 & 1203$\pm $16 & 1314 & 1340 \\
$^{28}$Al & 7.725 & $^{28}$Al+$^{12}$C & 19 & 1866$\pm $121 & 1732 & 1898 \\
$^{27}$P & 0.900 & $^{27}$P+$^{28}$Si & 34.5 & 2900$\pm $370 & 1972 & 2145
\\
$^{28}$P & 2.066 & $^{28}$P+$^{28}$Si & 39.0 & 2237$\pm $80 & 1934 & 2109 \\
\hline
\end{tabular}%
\thinspace
\end{table}%
\clearpage
\begin{center}
\begin{figure}[tbp]
\includegraphics[width=12cm,angle=0.]{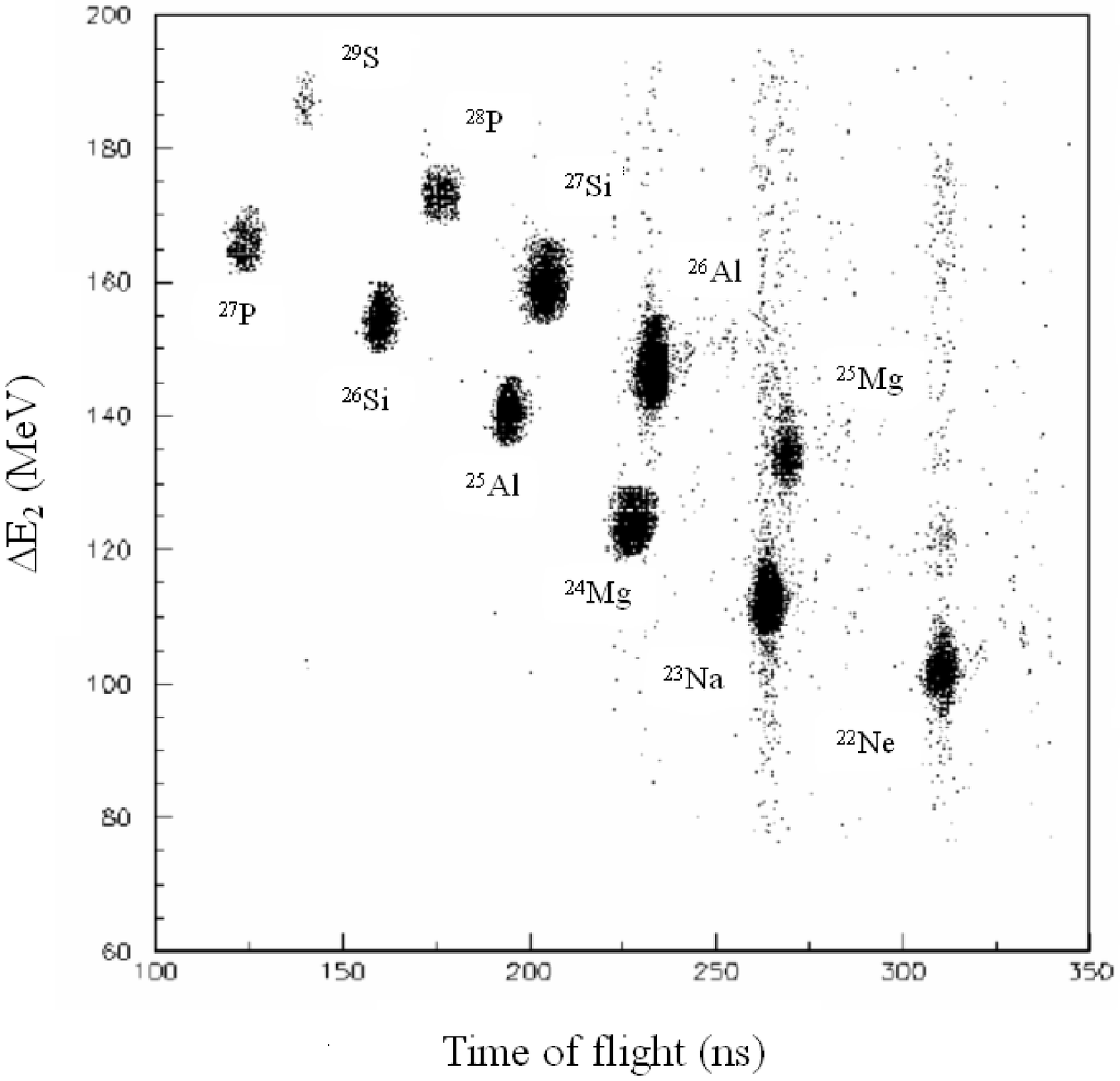}
\vspace*{0.5cm}
\caption{Two-dimensional plot of TOF versus $\Delta E_{2}$.}
\label{fig1}
\end{figure}

\clearpage
\begin{figure}[tbp]
\includegraphics[width=12cm,angle=-90.]{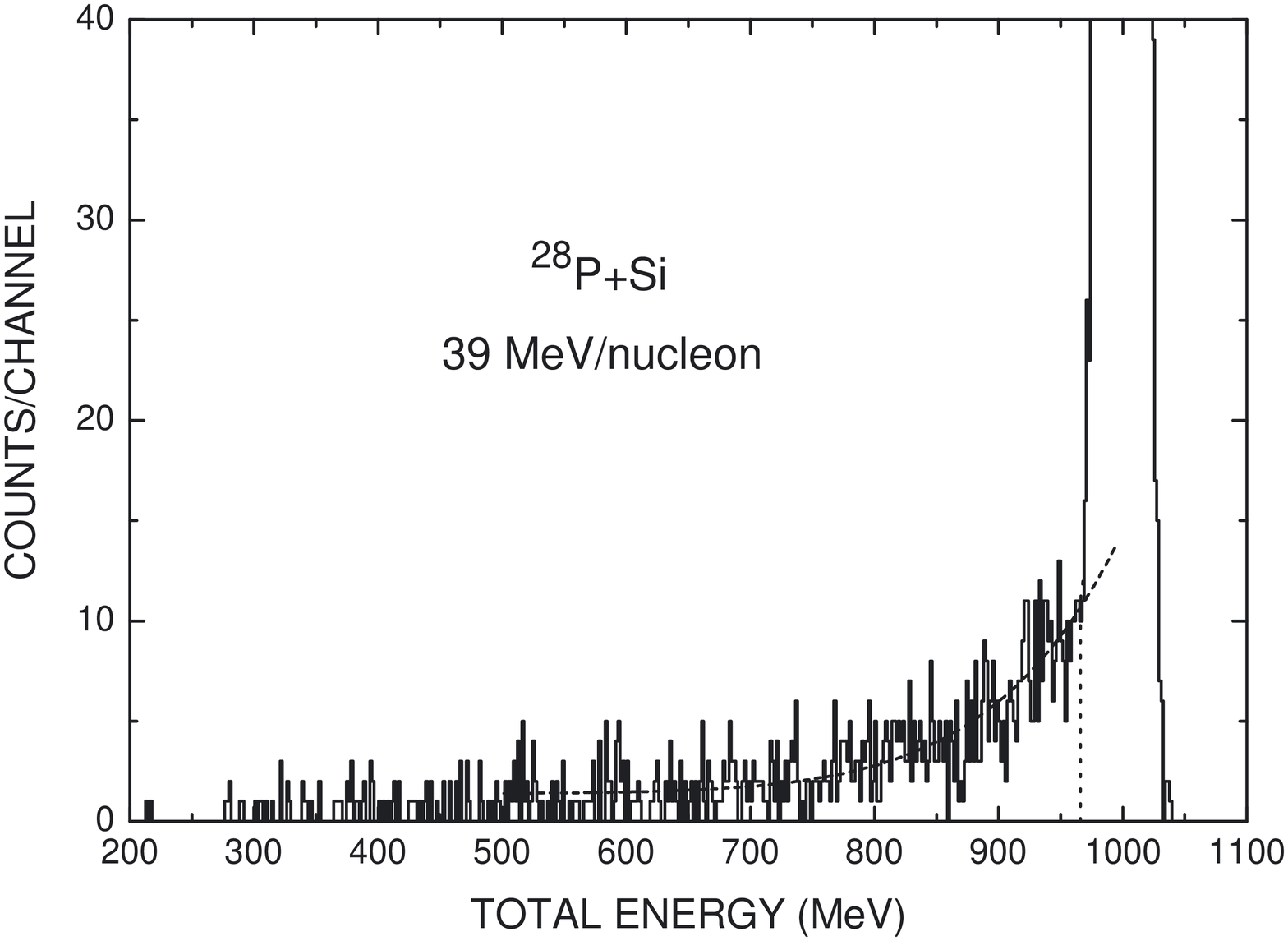} \vspace*{0.5cm}
\caption{Total energy deposition spectrum of $^{28}P$ projectile in Si
telescope. Events to the left of the dotted vertical line are counted as
reactions.}
\label{fig2}
\end{figure}
\clearpage
\begin{center}
\medskip

\begin{figure}[tbp]
\includegraphics[width=12cm,angle=0.]{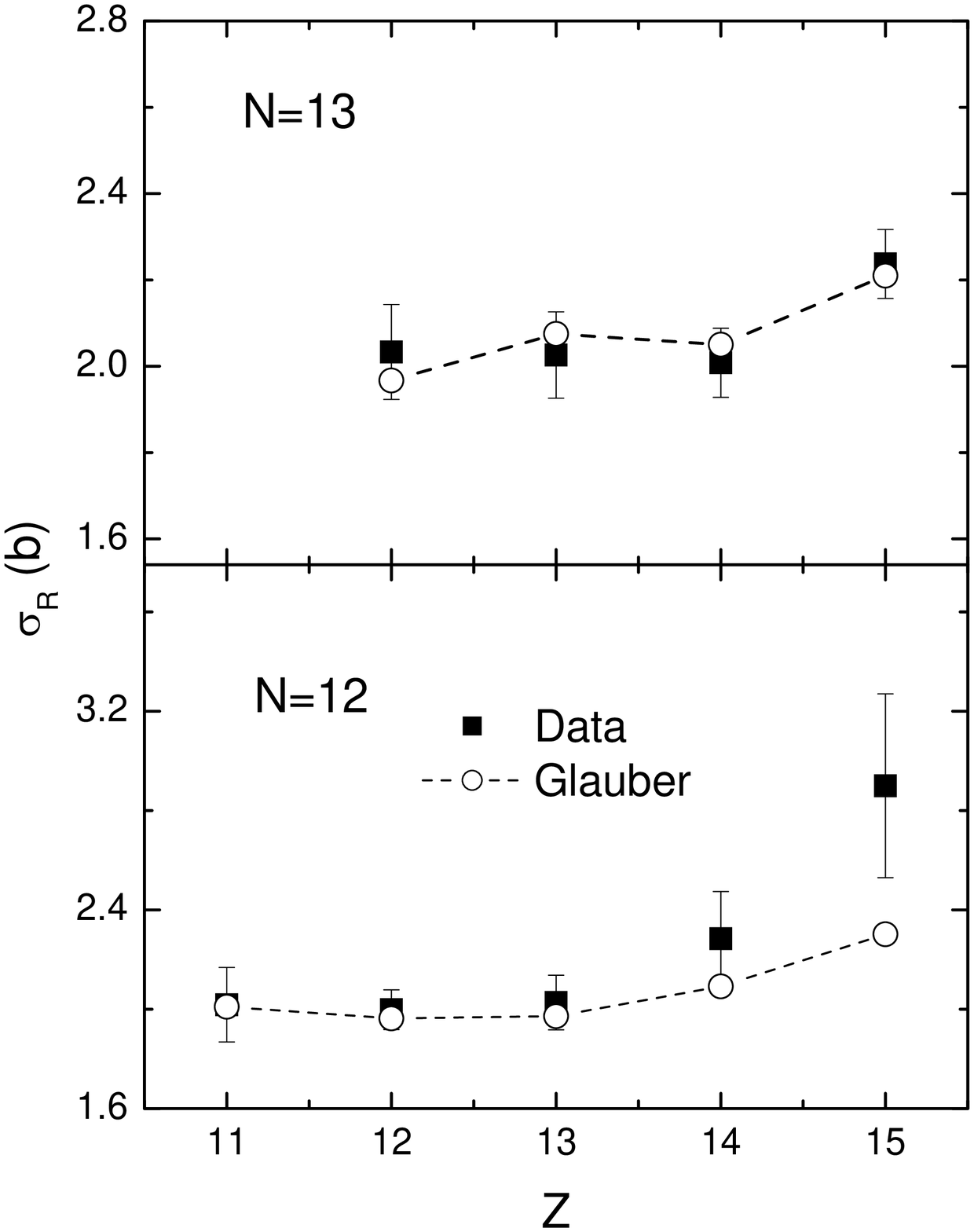} \vspace*{0.5cm}
\caption{The Z dependence of the reaction cross sections for the isotones
with N=12 and 13 at 40 MeV/nucleon. The solid squares with error bar
represent the experimental data. The open circles illustrate the prediction
of the modified Glauber model in the OL approach. The symbols are connected
by lines for each isotonic number to guide the eye.}
\label{fig_3}
\end{figure}

\clearpage\centering
\begin{figure}[tbp]
\includegraphics[width=12.cm,angle=-90.]{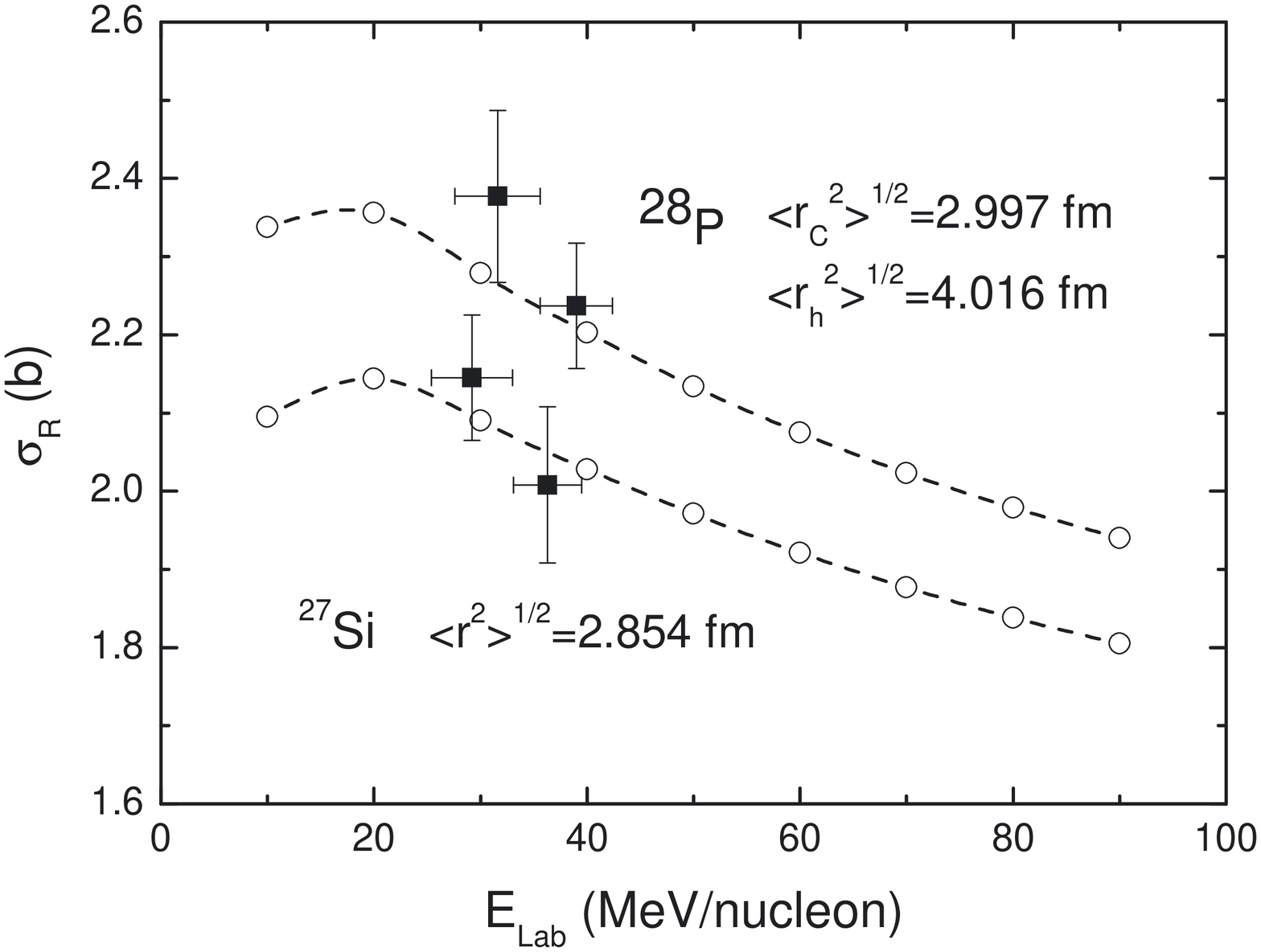} \vspace*{0.5cm}
\caption{Measured $\protect\sigma _{R}$ vs energy for the $^{27}$Si, $^{28}$%
P+$^{28}$Si reactions. The predictions of the modified Glauber model of the
OL approach (open circles) are compared with the experimental data.}
\label{fig_4}
\end{figure}

\begin{figure}[tbp]
\includegraphics[width=12cm,angle=0.]{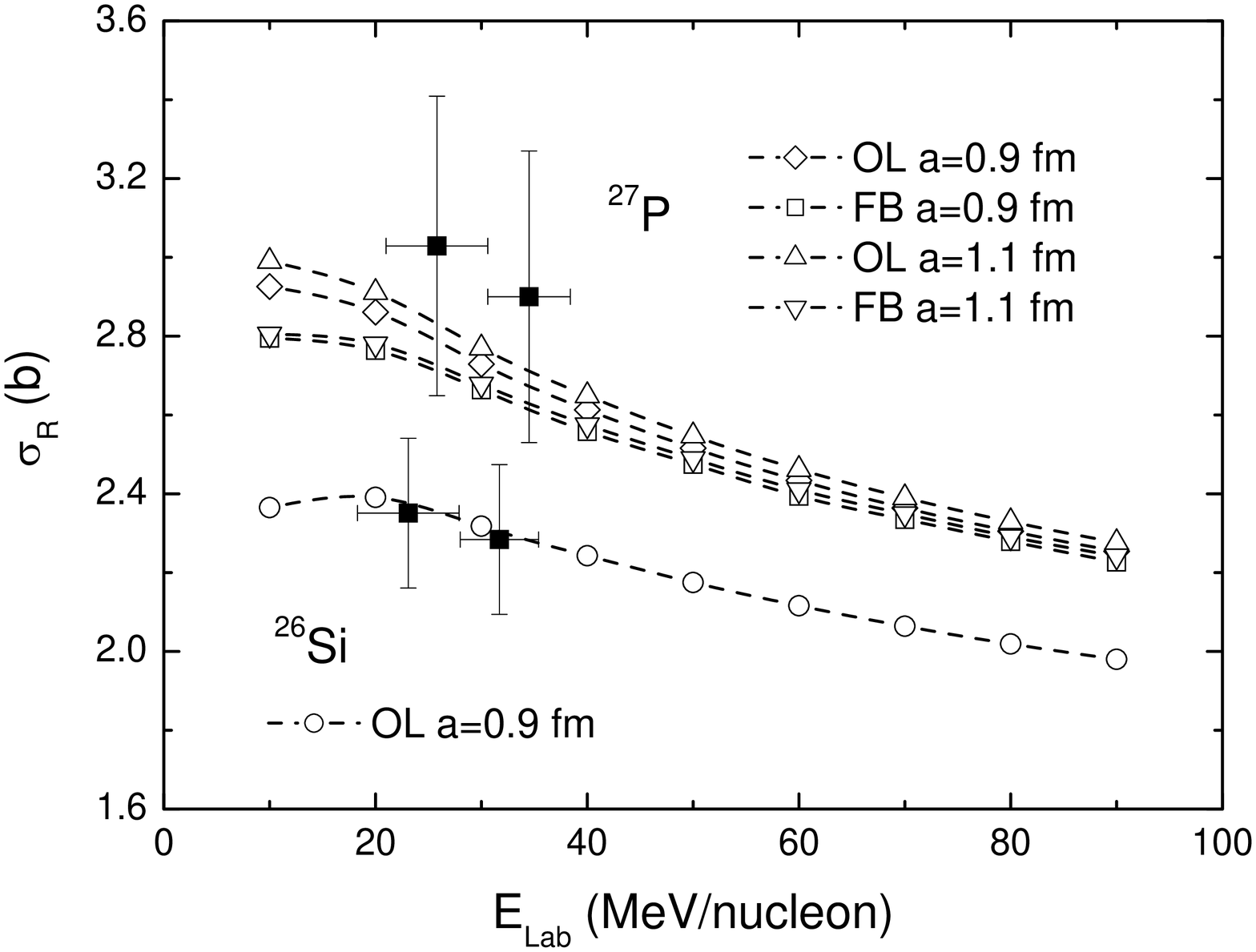} \vspace*{0.5cm}
\caption{Measured $\protect\sigma _{R}$ vs energy for the $^{26}$Si, $^{27}$%
P + $^{28}$Si reactions. The predictions of the modified Glauber model of OL
and FB approaches (open symbols) are compared with the experimental data.
The numbers in the figure represent the diffuseness parameter of the WS
potential for the valence proton.}
\label{fig_5}
\end{figure}
\end{center}
\end{center}
\end{document}